\documentclass[a4paper]{jpconf}
\usepackage{graphicx,color}
\usepackage{hyperref}
\usepackage{amsmath,amssymb}
\usepackage{subfig}
\usepackage[numbers,sort&compress,comma]{natbib}
\usepackage{cmap}
\hypersetup{colorlinks=true,citecolor=blue}

\begin{document}
\newcommand{\subfigureautorefname}{\figureautorefname}

\renewcommand{\equationautorefname}{Eq.}
\renewcommand{\figureautorefname}{Fig.}

\newcommand{\Int}{\int\limits}
\newcommand{\IInt}{\iint\limits}
\newcommand{\IIInt}{\iiint\limits}
\newcommand{\IIIInt}{\iiiint\limits}
\newcommand{\ui}{\mathrm{i}}
\newcommand{\ue}{\mathrm{e}}
\newcommand{\Vol}{\operatorname{vol}}
\newcommand{\R}{\mathbb{R}}
\newcommand{\Z}{\mathbb{Z}}
\newcommand{\N}{\mathbb{N}}
\newcommand{\cA}{\mathcal{A}}
\newcommand{\E}{\mathcal{E}}
\newcommand{\setsep}{ \;\; | \;\;}

\newcommand{\COMMENT}[1]{\textcolor{red}{\texttt{#1}}}
\newcommand{\ie}{i.e., }
\newcommand{\Schro}{Schr\"o\-din\-ger }
\newcommand{\eg}{e.g.\@ }
\newcommand{\cf}{cf.~}

\newcommand{\expval}[1]{\langle#1\rangle}
\newcommand{\abs}[1]{\left|#1\right|}
\newcommand{\norm}[1]{\left\|#1\right\|}
\newcommand{\bra}[1]{\langle#1|}
\newcommand{\ket}[1]{|#1\rangle}
\newcommand{\braket}[2]{\langle#1|#2\rangle}
\newcommand{\doubD}{{\mathord{\buildrel{\lower3pt\hbox{$\scriptscriptstyle\leftrightarrow$}}\over {\bf D}}}}
\newcommand{\unitv}[1]{\mathbf{\hat{#1}}}

\newcommand{\refeq}[1]{\hyperref[#1]{\equationautorefname~(\ref*{#1})}}

\newcommand{\cvec}[1]{\mathbf{#1}}
\newcommand{\op}[1]{\mathrm{\hat{#1}}}
\newcommand{\vecop}[1]{\cvec{\hat{#1}}}
\newcommand{\eqcomma}{\,,}
\newcommand{\eqstop}{\,.}
\newcommand{\ed}{\,}

\newcommand{\ddE}{\frac{\partial}{\partial E}}
\newcommand{\dx}{\dd x}
\newcommand{\dt}{\dd t}
\newcommand{\dr}{\dd r}
\newcommand{\dw}{\dd\omega}
\newcommand{\dwb}{\dd\bar{\omega}}
\newcommand{\dW}{\dd\Omega}
\newcommand{\dE}{\dd E}
\newcommand{\dk}{\dd k}
\newcommand{\dd}{\mathrm{d}}
\newcommand{\Dt}{\Delta t}
\newcommand{\hw}{\hbar\omega}

\newcommand{\sube}{{\mathrm{e}}}
\newcommand{\subn}{{\mathrm{n}}}

\newcommand{\csph}{{\mathcal{Y}}}
\newcommand{\sph}[2]{{\mathrm{Y}_{#2}^{(#1)}}}
\newcommand{\sphcmplx}[2]{{\mathrm{Y}_{#2}^{*(#1)}}}
\newcommand{\rensph}[2]{{\mathrm{C}_{#2}^{(#1)}}}
\newcommand{\renredsph}[1]{{\mathrm{C}^{(#1)}}}
\newcommand{\submax}{\mathrm{max}}
\newcommand{\Lmax}{L_\submax}
\newcommand{\lonemax}{l_{1,\submax}}
\newcommand{\ltwomax}{l_{2,\submax}}

\newcommand{\cmfs}{\,\mathrm{cm}^4\mathrm{s}}
\newcommand{\ev}{\,\mathrm{eV}}
\newcommand{\eV}{\ev}
\newcommand{\au}{\,\mathrm{a.u.}}
\newcommand{\nm}{\,\mathrm{nm}}
\newcommand{\Wcm}{\,\mathrm{W}/\mathrm{cm}^2}
\newcommand{\as}{\,\mathrm{as}}
\newcommand{\fs}{\,\mathrm{fs}}
\newcommand{\He}{\mathrm{He}}
\newcommand{\Hep}{\He^+}
\newcommand{\kone}{{\cvec{k}_1}}
\newcommand{\ktwo}{{\cvec{k}_2}}

\newcommand{\ti}{{t_\mathrm{(i)}}}
\newcommand{\tii}{{t_\mathrm{(ii)}}}
\newcommand{\tiii}{{t_\mathrm{(iii)}}}
\newcommand{\tcor}{{t_\mathrm{cor}}}
\newcommand{\Tp}{{T_{\mathrm{p}}}}
\newcommand{\Teff}{{T_{\mathrm{eff}}}}
\newcommand{\Tramp}{{T_{\mathrm{ramp}}}}
\newcommand{\Tfull}{{T_{\mathrm{full}}}}
\newcommand{\Ene}{{E_\mathrm{ne}}}
\newcommand{\Eeqs}{{E_\mathrm{eq}}}
\newcommand{\Eexc}{{E_\mathrm{exc}}}
\newcommand{\Etot}{{E_\mathrm{tot}}}
\newcommand{\PDI}{{P^{DI}}}
\newcommand{\PDIs}{{P^{DI}_\mathrm{seq}}}
\newcommand{\PDIns}{{P^{DI}_\mathrm{nonseq}}}
\newcommand{\Pasym}{\mathcal{A}}
\newcommand{\DE}{\Delta E}
\newcommand{\G}{\mathcal{G}}
\newcommand{\level}[3]{{}^{#1}{#2}^{\textrm{#3}}}
\newcommand{\redmate}[3]{\left<#1\left|\left|#2\right|\right|#3\right>}

\title{Probing scattering phase shifts by attosecond streaking}

\author{R Pazourek$^{1}$,
S~Nagele$^{1}$,
K~Doblhoff-Dier$^{1}$,
J~Feist$^{2}$,
C~Lemell$^{1}$,
K~T\H{o}k\'{e}si$^{3}$ and
J~Burgd\"orfer$^{1}$}

\address{$^1$ Institute for Theoretical Physics, Vienna University of Technology, 1040 Vienna, Austria, EU}
\address{$^2$ ITAMP, Harvard-Smithsonian Center for Astrophysics, Cambridge, Massachusetts 02138, USA}
\address{$^3$ Institute of Nuclear Research of the Hungarian Academy of Sciences (ATOMKI), 4001 Debrecen, Hungary, EU}

\ead{renate.pazourek@tuwien.ac.at}

\begin{abstract}
Attosecond streaking is one of the most fundamental processes in attosecond science allowing for a mapping of temporal (i.e.\ phase) information on the energy domain.
We show that on the single-particle level attosecond streaking time shifts contain spectral phase information associated with the Eisenbud-Wigner-Smith (EWS) time delay, provided the influence of the streaking infrared field is properly accounted for. 
While the streaking phase shifts for short-ranged potentials agree with the associated EWS delays, Coulomb potentials require special care. 
We show that the interaction between the outgoing electron and the combined Coulomb and IR laser fields lead to a streaking phase shift that can be described classically.

\end{abstract}

\section{Introduction}

The emerging field of \emph{attoscience} \cite{HenKieSpi2001, DreHenKie2001, ItaQueYud2002, KieGouUib2004, YakBamScr2005, QueMaiIta2005, SanBenCal2006} enables the investigation of electron dynamics as well as timing information of photoionization processes. 
Attosecond streaking has developed into a powerful tool to achieve temporal-resolution on the sub-100 attosecond time scale~\cite{DreHenKie2002, GouUibKie2004, CavMueUph2007,SchFieKar2010}. It is based on a pump-probe setting with an extreme ultraviolet (XUV) pulse of a few hundred attoseconds duration serving as pump and a phase-controlled few-cycle infrared (IR) pulse as probe.
Temporal information about the photoionization process can thus be mapped onto the energy axis in analogy to conventional streaking.
First proof-of-principle implementations enabled the direct measurement of the life time of the Kr($3d^{-1}$) hole by Auger decay \cite{DreHenKie2002}, the first measurement of the time-dependent electric field of an IR light wave \cite{GouUibKie2004}; as well as measurements of delayed photoemission from core levels relative to conduction band states of a tungsten surface \cite{CavMueUph2007} and from different shells of atomic neon \cite{SchFieKar2010}. In the latter case, the measured streaking time shift is one order of magnitude shorter than the XUV ``pump'' pulse and two orders shorter than the oscillation period $T$ of the IR probe pulse.

The challenge in interpreting the obtained time shifts lies in disentangling the modifications of the observed streaking time shifts by the IR probe field from the intrinsic time shifts one is interested in. These intrinsic time delays are connected to the spectral variation of the scattering phases by the Eisenbud-Wigner-Smith (EWS) time delay operator~\cite{Wig1955,Smi1960}. 
In this contribution we will apply the EWS time delay formalism to photoionization (half-scattering) and show with the help of numerical simulations that the EWS time delay and thus scattering phases become accessible by attosecond streaking as long as the potential is short-ranged.
In particular, we study attosecond streaking of the release time of electrons in atomic photoemission~\cite{SchFieKar2010, NagPazFei2011} by solving the time-dependent Schr\"odinger equation for effective one-electron systems. 
While the EWS time shift only contains the scattering phase from the single-photon dipole transition, in streaking setups an additional (non-perturbative, in contrast to RABITT, \cf \cite{KluDahGis2011}) IR field is present.
To disentangle effects that stem from distortions of the initial state by the IR field we employ a ``restricted ionization model'' \cite{Schafer2009, MauJohMan2008}. 
In addition, we perform classical-trajectory Monte Carlo simulations to account for trajectory effects on the time shift resulting from the interaction between the outgoing electron and the combined Coulomb and IR laser fields. 
When both distortion effects are accounted for, the EWS time shift agrees well with the associated streaking time shift for short-ranged potentials which we will show for the example of the centrifugal potential. 
Unless otherwise stated, atomic units are used.

\section{Computational Methods}
 
The Schr\"odinger equation for a one-electron system in the presence of an external IR and XUV field is given by
\begin{equation}
i\frac{\partial}{\partial t}\vert\psi\rangle=\hat{H}\vert\psi\rangle \, , \quad 
\hat{H} = \hat{H}_\mathrm{a}+\hat{H}_{\scriptscriptstyle{\scriptscriptstyle\mathrm{IR}}}+\hat{H}_{\scriptscriptstyle\mathrm{XUV}} \, ,
\label{eq:schrodinger}
\end{equation} 
where $\hat{H}_a$ is the atomic Hamiltonian with an effective one-electron potential, $\hat{H}_{\scriptscriptstyle\mathrm{XUV}}=\vec r\cdot\vec{F}_{\scriptscriptstyle\mathrm{XUV}}(t)$ describes the interaction with the the electric field $\vec{F}_{\scriptscriptstyle\mathrm{XUV}}$ of the attosecond pump pulse and $\hat{H}_{\scriptscriptstyle\mathrm{IR}}=\vec r\cdot\vec{F}_{\scriptscriptstyle\mathrm{IR}}(t)$ the interaction with the IR probe pulse. Typical energies and intensities of the electric fields are $\hbar\omega_{\scriptscriptstyle\mathrm{XUV}}\approx 100\ev$ and $I_{\scriptscriptstyle\mathrm{XUV}}\leq 10^{13}\Wcm$ for the attosecond pump pulse, and $\hbar\omega_{\scriptscriptstyle\mathrm{IR}}\approx 1.5\ev$ and $I_{\scriptscriptstyle\mathrm{IR}}\approx 10^{10}-10^{12}\Wcm$ for the probe field. 

Our computational method for solving \autoref{eq:schrodinger} is based on the well-established pseudo-spectral split-operator method as described in \cite{TonChu1997}. 
We also employ a constrained Schr\"odinger equation, in which the interaction of the bound state with the probe field is suppressed. This so-called ``restricted ionization model'' (RIM) \cite{Schafer2009,MauJohMan2008} is based on the decomposition of the wave function $\psi(\vec{r}, t)$ into an initial state part $\Phi_0(\vec{r})$ and a continuum part $\tilde{\Phi}(\vec{r}, t)$
\begin{equation}
\vert\psi(t)\rangle=e^{-iE_0t}\vert\Phi_0\rangle+e^{i\tilde{E}t}\vert \tilde{\Phi}(t) \rangle,
\label{eq:rimAnsatz}
\end{equation}
where $E_0 \Phi_0(\vec{r})=\hat{H}_\mathrm{a}(\vec{r})\Phi_0(\vec{r})$ and $\tilde{E}=E_0+\omega_{\scriptscriptstyle\mathrm{XUV}}$. Inserting this ansatz into the Schr\"odinger equation \eqref{eq:schrodinger}, leads to the following (exact) expression describing the time evolution of the continuum wave function $\tilde{\Phi}(\vec{r}, t)$,
\begin{equation}
	i\frac{\partial}{\partial t}\vert \tilde{\Phi}(t) \rangle=	\left(\hat{H}_\mathrm{a}+\hat{H}_{\scriptscriptstyle\mathrm{IR}}+\hat{H}_{\scriptscriptstyle\mathrm{XUV}}-\tilde{E}\right)\vert \tilde{\Phi}(t) \rangle+	\left(e^{i\omega_{\scriptscriptstyle\mathrm{XUV}}t}\hat{H}_{\scriptscriptstyle\mathrm{IR}}+\hat{H}_{\scriptscriptstyle\mathrm{XUV}}^{+}+e^{2i\omega_{\scriptscriptstyle\mathrm{XUV}}t}\hat{H}_{\scriptscriptstyle\mathrm{XUV}}^{-}\right)\vert\Phi_0\rangle,
	\label{eq:rimSchrodingerFull}
\end{equation}
where $\hat{H}_{\scriptscriptstyle\mathrm{XUV}}=e^{i\omega_{\scriptscriptstyle\mathrm{XUV}}t}\hat{H}_{\scriptscriptstyle\mathrm{XUV}}^{-}+e^{-i\omega_{\scriptscriptstyle\mathrm{XUV}}t}\hat{H}_{\scriptscriptstyle\mathrm{XUV}}^{+}$. Following Schafer \emph {et al.}~\cite{Schafer2009}, we apply the following simplifications:
We neglect the interaction of the XUV pulse with the continuum wave packet and drop the highly oscillatory terms $e^{2i\omega_{\scriptscriptstyle\mathrm{XUV}}t}\hat{H}_{\scriptscriptstyle\mathrm{XUV}}^{-}$ and $e^{i\omega_{\scriptscriptstyle\mathrm{XUV}}t}\hat{H_{\scriptscriptstyle\mathrm{IR}}}$ acting on the initial state (slowly varying envelope approximation).
The latter simplification suppresses the interaction of the IR pulse with the initial state, thereby inhibiting ground-state polarization and excitation. Their influence on the streaking time shifts can thus be investigated by comparing RIM and full TDSE results. 
The equation governing the time evolution of the continuum wave function then has the structure of a Schr\"odinger equation with an additional source term and can hence be solved in the same way as \autoref{eq:schrodinger}.

\section{Results}
We will start our discussion on the accessibility of scattering phases by attosecond streaking with an investigation of streaking time shifts for atomic photoionization in hydrogenic systems.
The attosecond streaking technique is based on the assumption that the asymptotic momentum of an electron ionized by an ultrashort (few hundred attoseconds) XUV pulse is shifted by the instantaneous value of the vector potential $\vec A_{{\scriptscriptstyle\mathrm{IR}}}(\tau)$ of the IR streaking field  at the moment of ionization,
\begin {equation}
\label{eq:streaking_simple}
\vec p_f(\tau)=\vec p_0 - \vec A_{{\scriptscriptstyle\mathrm{IR}}}(\tau) \eqstop
\end{equation}
Here, $p_0 = [2(\omega_{{\scriptscriptstyle\mathrm{XUV}}}-\mathcal{E}_i)]^\frac{1}{2}$ is the unperturbed asymptotic photoelectron momentum, where $\omega_{{\scriptscriptstyle\mathrm{XUV}}}$ is the XUV photon energy and $\mathcal{E}_i$ the initial binding energy (all in a.u.).
For temporally well-defined IR electric streaking fields $\vec F_{{\scriptscriptstyle\mathrm{IR}}}(t)$ and vector potentials
$\vec A_{{\scriptscriptstyle\mathrm{IR}}}(\tau) = - \int_\tau^\infty  \vec F_{{\scriptscriptstyle\mathrm{IR}}}(t) dt$
the momentum shift (energy shift) of the photoelectron allows thus, in principle, a mapping of emission time $\tau$ onto energy (`streaking'). 

\begin{figure}[tb]
  \centering
  \subfloat[][]{
  \includegraphics[width=0.45\linewidth]{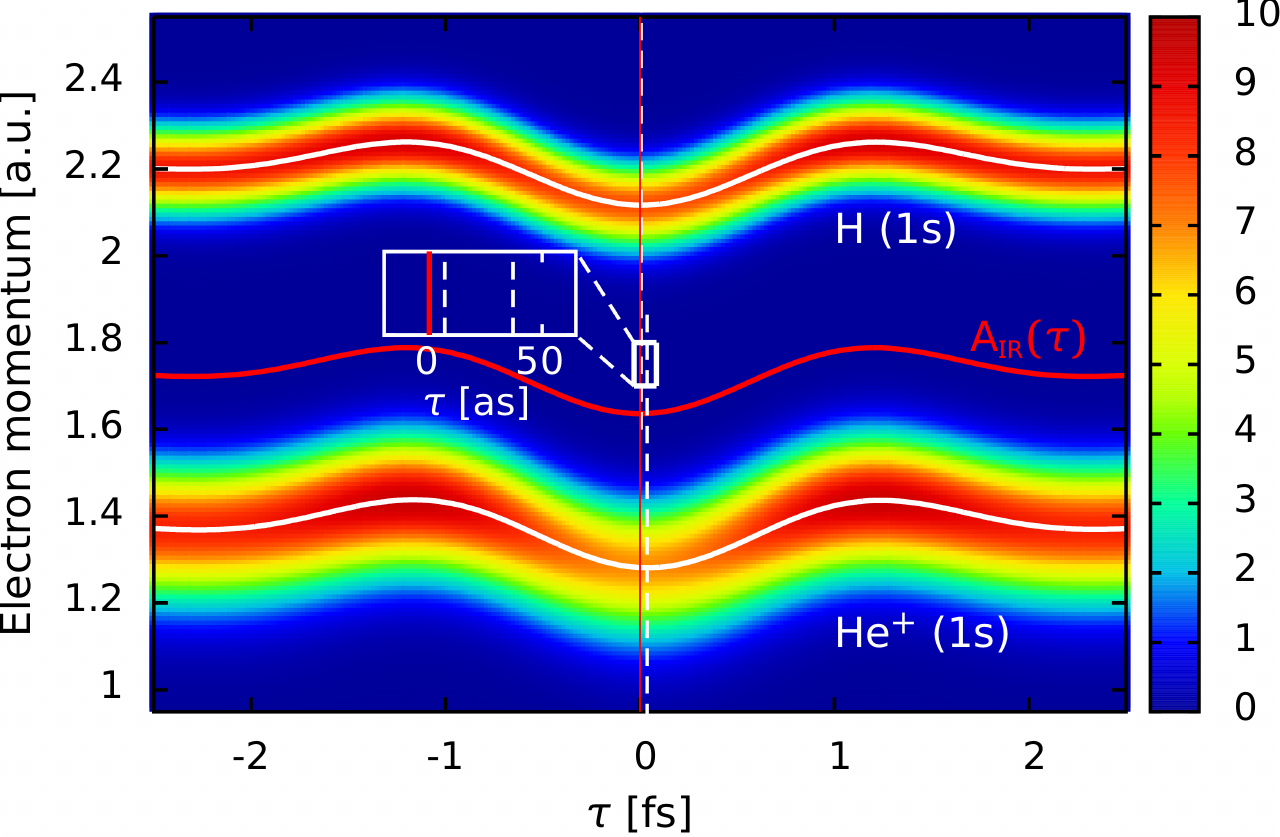}\label{fig:spectrogram}}
  \subfloat[][]{
  \includegraphics[width=0.45\linewidth]{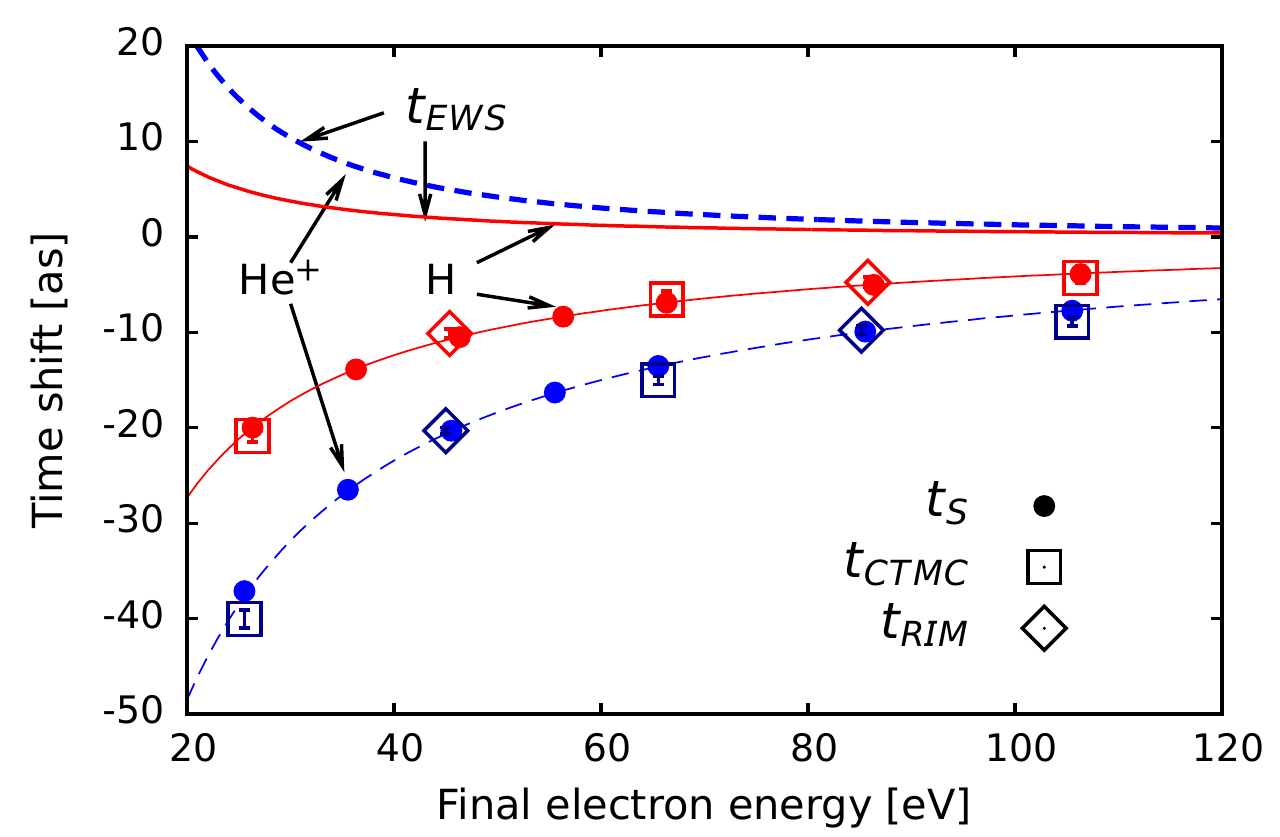} \label{fig:timeshifts}}
  \caption{(a) Streaking spectrograms for an 800 nm IR laser field with a duration of 3\,fs and an intensity of $10^{12}$\,W/cm$^2$. The graphs show the final momentum distribution along the laser polarization axis for a H($1s$) and $\Hep$($1s$) initial state and XUV photon energies of $80\ev$ together with the vector potential $A_{\scriptscriptstyle\mathrm{IR}}(\tau)$. The solid white lines are the first moments of the electron spectra. The dashed white vertical lines indicate the shift of the central minimum relative to the vector potential (red solid line). 
  (b) Temporal shifts $t_S$ extracted from quantum mechanical streaking simulations with the full TDSE (full circles) and RIM (open diamonds), classical (CTMC) streaking simulations (open squares), and for comparison, the EWS time shift $t_\mathrm{EWS} = d\varphi /dE$ applied to the Coulomb phase (H$^+$: red solid line, He$^{2+}$: blue dashed line).}
\end{figure}

\autoref{fig:spectrogram} shows photoionization spectrograms for ionization from the hydrogen and $\Hep$ ground state by an $80\ev$ XUV pulse with a duration of 200\,as (FWHM of the Gaussian intensity envelope), streaked by an 800\,nm IR laser field with a duration of 3\,fs (FWHM of the $\cos^2$ envelope) and an intensity of $10^{12}$\,W/cm$^2$. 
By fitting the streaking curves (first moment of the final electron momentum distribution, white lines in \autoref{fig:spectrogram}) to the analytic form of the IR vector potential $\vec A_{{\scriptscriptstyle\mathrm{IR}}}(t+t_S)$, we obtain the absolute delays shown in \autoref{fig:timeshifts}. 
All results are obtained by a nonlinear least-squares fit (error bars are present in all figures) of the first moment of the spectrogram taken along the laser polarization axis with an opening angle of 10$^\circ$.
The spectrograms are obtained with a full numerical solution of the 3D TDSE without any approximations. 
The sign convention for $t_S$ ensures that \emph{positive} values correspond to \emph{delayed} emission, i.e.\ the electron `feels' the vector potential at a \emph{later} time. 
A negative time shift, or time advance, $t_S < 0$, relative to the vector potential can be observed over a wide range of final electron energies (\autoref{fig:timeshifts}) corresponding to varying the XUV pulse frequency $\omega_{{\scriptscriptstyle\mathrm{XUV}}}$. 

The question arises whether these time advances are related to the intrinsic quantum mechanical time shift corresponding to the scattering phases of the system.
For this intrinsic time shift we use the single channel form of the Eisenbud-Wigner-Smith time shift $t_{\scriptscriptstyle\mathrm{EWS}}$ \cite{Wig1955, Smi1960} given by the energy derivative of the spectral phase $\varphi=\arg(\bra{\vec p_0}z\ket{\psi_i})$, i.e.\ the group delay \cite{deCNus2002} of the electronic wave packet
\begin{equation}
\label{eq:group_delay}
t_{\scriptscriptstyle\mathrm{EWS}} = \frac{d\varphi}{dE}.
\end{equation}
For photoionization in pure Coulomb potentials the phase $\varphi$ is given by the Coulomb phase
$\sigma_\ell = \arg[\Gamma(\ell+1-iZ/k)]$, 
where $\ell$ is the angular momentum of the free electron, $k$ its wavenumber, and $Z$ is the charge of the remaining ion. However, the EWS time shift is only correctly defined for short-ranged potentials, so it comes as no great surprise that $d\sigma_\ell/dE$ does not agree with the time shift extracted from streaking (\cf\autoref{fig:timeshifts}). 

The physical interpretation of the extracted streaking time shifts requires a detailed inquiry into the assumptions underlying \autoref{eq:streaking_simple}. 
The fundamental assumption of the streaking field as a probe is that the IR field does not distort the system under scrutiny. 
In the following we investigate the influence of the IR field on the entrance and exit channels. 
The original attosecond streaking model \cite{ItaQueYud2002} is based on the strong-field approximation which assumes a `sudden' transition to the undisturbed momentum $\vec p_0$ caused by photoabsorption (\autoref{eq:streaking_simple}) followed by a momentum shift by the IR field of the remaining streaking pulse.
However, as the wave packet recedes from the ionic core, it propagates in the atomic (ionic) potential with a local momentum $\vec p(\vec r)$ rather than the asymptotic momentum $\vec p_0$. 
Thus, neglect of the deviation of the local from the asymptotic momentum is no longer valid for streaking in the Coulomb field (\cf \cite{SmiSpaIva2006a}). 
Our classical trajectory Monte Carlo (CTMC) simulation (see \cite{NagPazFei2011} for details) gives streaking time shifts in almost perfect agreement with the full TDSE over a wide range of final-state energies for both H$(1s)$ and He$^+(1s)$ initial states (\autoref{fig:timeshifts}) and over two orders of magnitude in intensity ($I_{\scriptscriptstyle\mathrm{IR}} \approx 10^{10} - 10^{12}\Wcm$) of the streaking field (not shown). We therefore conclude that the energy dependence of the streaking time shifts results from classical Coulomb-laser coupling \cite{NagPazFei2011} (see also \cite{SmiSpaIva2006a}). This has also been identified in one-dimensional studies using the eikonal approximation \cite{ZhaThu2010} and by an alternative interferometric method for extraction of time shifts and atomic phases (`RABITT') \cite{KluDahGis2011}.

\begin{figure}[tb]
  \centering
  \subfloat[][TDSE]{
    \includegraphics[width=0.48\linewidth]{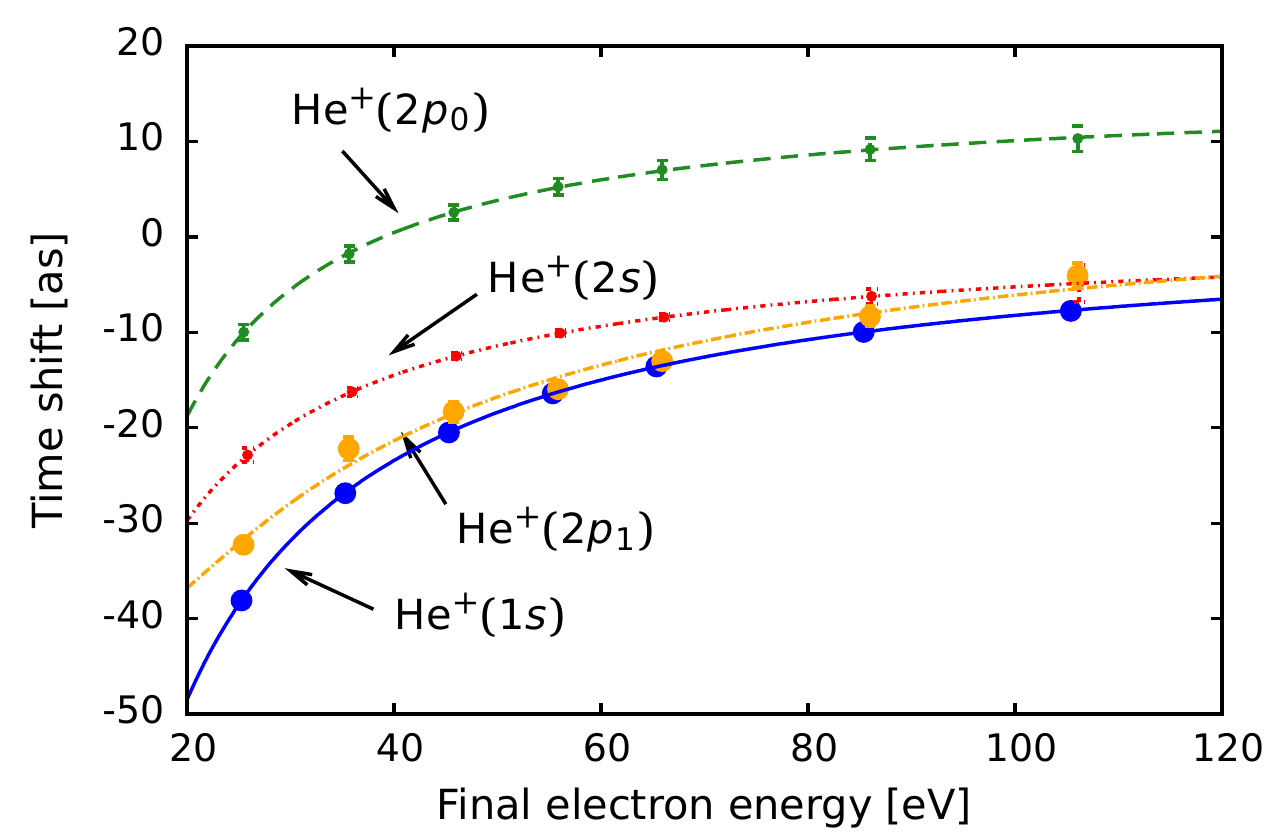}\label{fig:timeshifts_l_tdse}}
  \subfloat[][RIM]{
    \includegraphics[width=0.48\linewidth]{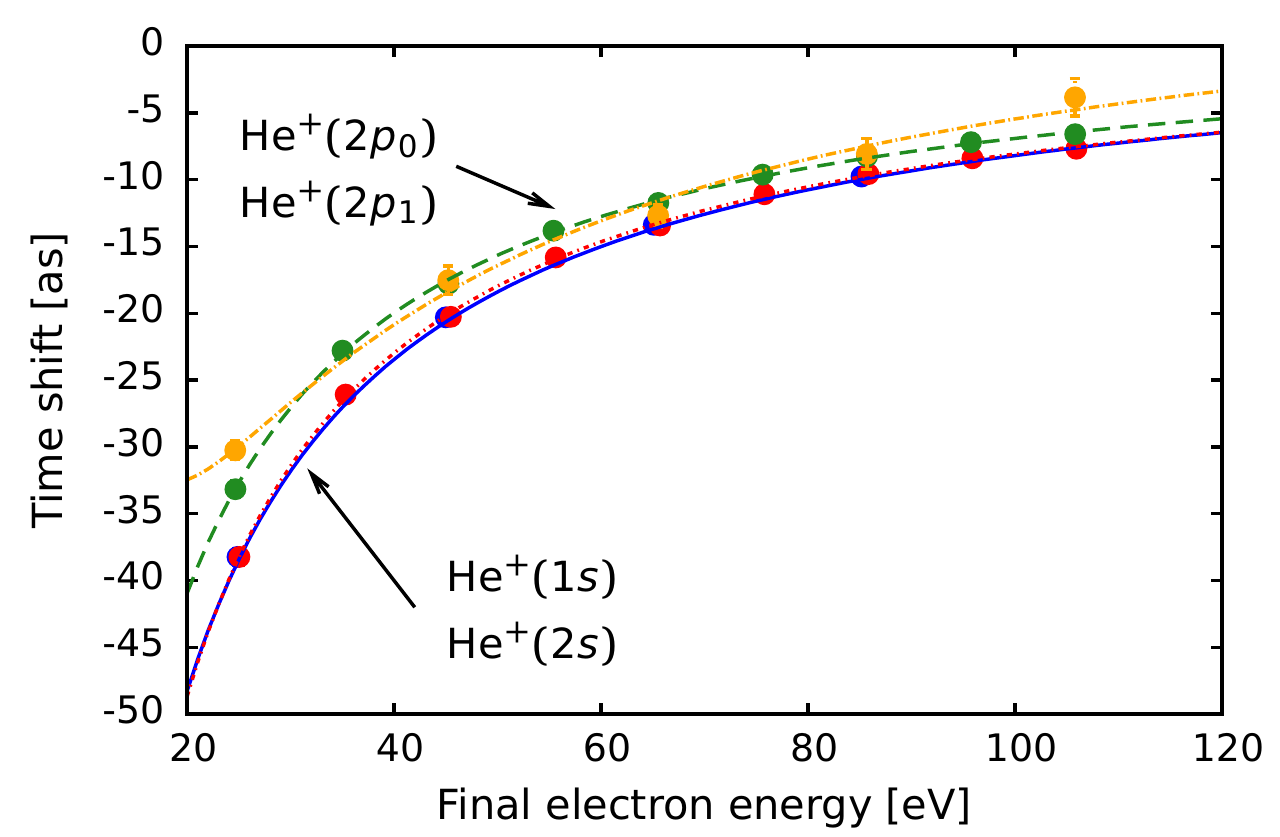}\label{fig:timeshift_l_rim}}  
  \caption{Streaking time shifts $t_S$ extracted from quantum mechanical streaking simulations with full TDSE (a) and RIM (b) for different initial states. The remaining difference between initial s and p states in the RIM results (b) can be attributed to the short-ranged scattering phase shift induced by the centrifugal potential (see \autoref{fig:timeshift_l_cent}). 
 }
  \label{fig:timeshifts_l}
\end{figure}

Initial state distortions by the IR field can be investigated by the constrained TDSE calculations (RIM) where the IR field explicitly only acts on the ionized part of the wave function. 
For energetically isolated initial states with small polarizability (H(1s), $\Hep$(1s),...) the RIM results agree perfectly with the full TDSE results (see \autoref{fig:timeshifts}), therefore excluding initial-state polarization effects. 
However, if the initial state is not isolated, initial-state polarization can strongly affect the extracted delays. 
In \autoref{fig:timeshifts_l} we show TDSE and RIM streaking results for low-lying states of He$^+(n\ell m)$ which are stable against tunnel ionization by the IR field ($I_{\scriptscriptstyle\mathrm{IR}} \lesssim 10^{12}\Wcm$). 
Within the full TDSE calculation we find remarkably strong initial-state dependent streaking field distortions with relative time shifts of up to $20\as$ between states with $n=2$ and different quantum numbers\footnote{The strong variation of $t_S$ with angular momentum $\ell$ is not specific to the degeneracy of the hydrogenic He$^+(n=2)$ manifold. By choosing an atomic model potential that breaks the $SO(4)$ symmetry we have verified that the streaking time shift persists for non-hydrogenic $\ell$ manifolds, thereby excluding degeneracy effects as the origin. Moreover, transient inter-shell coupling was found to dominate dynamical polarization over intra-shell mixing.} $\ell$ and $m$. 
This pronounced streaking phase shift as a function of angular momentum differs from recent calculations \cite{BagMad2010Err,ZhaThu2010} which did not find a shift for states with (initially) vanishing dipole moment.
However, for the simulations using the RIM the relative time shifts between 1s and 2s disappear completely, and the time shifts for $2p_0$ and $2p_1$ almost agree (note that for initial $p$ states the angular distributions are more complex and $t_S$ becomes also more sensitive to the opening angle.)  
We therefore can rule out a \emph{state-dependent} exit channel distortion of the IR field.
 
\begin{figure}[tb]
  \centering
  \begin{minipage}{0.49\linewidth}
  \includegraphics[width=0.95\linewidth]{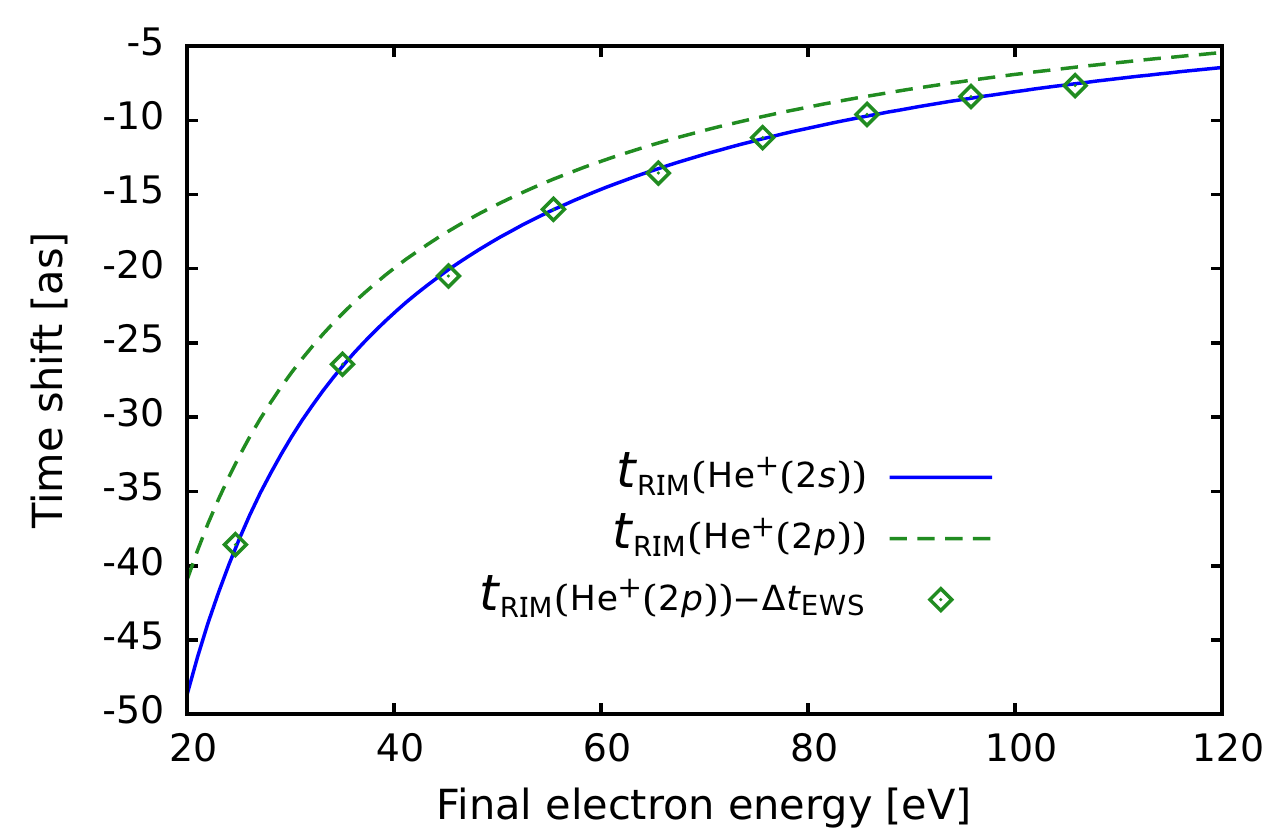}
  \end{minipage}
  \begin{minipage}{0.49\linewidth}
  \caption  {Streaking time shifts $t_{\scriptscriptstyle\mathrm{RIM}}$ (lines) for excited initial states ($2s$ and $2p_0$) in $\Hep$ (see               \autoref{fig:timeshift_l_rim}). The difference between $t_{\scriptscriptstyle\mathrm{RIM}}(2p_0)$ and $t_{\scriptscriptstyle\mathrm{RIM}}(2s)$ is given by $\Delta t_{\scriptscriptstyle\mathrm{EWS}}=\ddE \arg(\bra{\vec p_0}z\ket{\phi_{2p_0}}-\ddE \arg(\bra{\vec p_0}z\ket{\phi_{2s}})$, which is given by the different centrifugal potential seen by the outgoing electron ($V_l=l(l+1)/(2r^2)$ with $l$ being the angular momentum of the free electron, $l=p$ for initial $s$ states and $l=s$ or $l=d$ for initial $p$ states).
  }\label{fig:timeshift_l_cent}
  \end{minipage}
\end{figure}

The remaining difference between $s$ and $p$ states, after we have removed initial state distortion effects, can be finally attributed to the EWS time shift $\Delta t_{\scriptscriptstyle\mathrm{EWS}}=\ddE \arg(\bra{\vec p_0}z\ket{\phi_{2p_0}}-\ddE \arg(\bra{\vec p_0}z\ket{\phi_{2s}})$ as is shown in \autoref{fig:timeshift_l_cent}. 
Here $\ddE \arg(\bra{\vec p_0}z\ket{\phi_{2p_0}}$ still includes the Coulomb phase shift $\sigma_l$, however, for different angular momenta $l$ of the final state. 
The effective potential includes the long range Coulomb potential $-Z/r$, which is the same for both initial states, and the different centrifugal potential $V_l=l(l+1)/(2r^2)$.
Thus we find 
that $1/r^2$ potentials are sufficiently short-ranged so that the corresponding scattering phases are accessible by streaking.

We have shown in \cite{NagPazFei2011} that for pure short-ranged Yukawa potentials as well as for a short-ranged admixture to the Coulomb potential
, the EWS-delay associated with the short-range part of the scattering phase agrees remarkably well with the observed streaking time shift.  
Along the same lines it is possible to describe streaking time shifts in static atomic model potentials by the Coulomb-laser coupling time shift of the asymptotic Coulomb field and the EWS time delay of the remaining short-ranged parts of the model potential as long as initial-state polarization can be neglected.


\section{Summary}
We have shown that Eisenbud-Wigner-Smith (EWS) time shifts (or energy variation of the scattering phase) for short-ranged potentials become accessible by attosecond streaking provided both initial-state dependent entrance channel and final-state exit channel distortions are properly accounted for.
For Coulomb potentials the coupling between the IR streaking field and the Coulomb field which depends on the final energy of the free electron dominates the extracted streaking time shift but can be accounted for classically.
In addition we identified considerable state-dependent time shifts for easily polarizable initial states which are of quantum mechanical origin. 
Accounting for polarization of the initial state, the remaining difference of time delays between ionization from states with different angular momentum can be related to the EWS delay of the centrifugal potential. \\

This work was supported by the FWF-Austria, Grants No.\ SFB016 and P21141-N16, the TeT under Grant No.\ AT-2/2009, and in part by the National Science Foundation through TeraGrid resources provided by NICS and TACC under Grant TG-PHY090031. The computational results presented have also been achieved in part using the Vienna Scientific Cluster (VSC). JF acknowledges support by the NSF through a grant to ITAMP.


\bibliographystyle{jpbjo}
\bibliography{citeulike_atto_cleaned}

\providecommand{\newblock}{}
\begin{thebibliography}{10}
\providecommand{\url}[1]{\texttt{#1}}
\providecommand{\urlprefix}{URL }
\providecommand{\eprint}[2][]{\url{#2}}

\bibitem{HenKieSpi2001}
Hentschel M, Kienberger R, Spielmann C, Reider G~A \emph{et~al.} 2001
  \emph{Nature} \textbf{414} 509

\bibitem{DreHenKie2001}
Drescher M, Hentschel M, Kienberger R, Tempea G \emph{et~al.} 2001
  \emph{Science} \textbf{291} 1923

\bibitem{ItaQueYud2002}
Itatani J, Qu\'{e}r\'{e} F, Yudin G~L, Ivanov, Krausz F and Corkum P~B 2002
  \emph{Phys. Rev. Lett.} \textbf{88} 173903

\bibitem{KieGouUib2004}
Kienberger R, Goulielmakis E, Uiberacker M, Baltuska A \emph{et~al.} 2004
  \emph{Nature} \textbf{427} 817

\bibitem{YakBamScr2005}
Yakovlev V~S, Bammer F and Scrinzi A 2005 \emph{J. Mod. Opt.} \textbf{52} 395

\bibitem{QueMaiIta2005}
Qu\'{e}r\'{e} F, Mairesse Y and Itatani J 2005 \emph{J. Mod. Opt.} \textbf{52}
  339

\bibitem{SanBenCal2006}
Sansone G, Benedetti E, Calegari F, Vozzi C \emph{et~al.} 2006 \emph{Science}
  \textbf{314} 443

\bibitem{DreHenKie2002}
Drescher M, Hentschel M, Kienberger R, Uiberacker M \emph{et~al.} 2002
  \emph{Nature} \textbf{419} 803

\bibitem{GouUibKie2004}
Goulielmakis E, Uiberacker M, Kienberger R, Baltuska A \emph{et~al.} 2004
  \emph{Science} \textbf{305} 1267

\bibitem{CavMueUph2007}
Cavalieri A~L, M\"{u}ller N, Uphues T, Yakovlev V~S \emph{et~al.} 2007
  \emph{Nature} \textbf{449} 1029

\bibitem{SchFieKar2010}
Schultze M, Fiess M, Karpowicz N, Gagnon J \emph{et~al.} 2010 \emph{Science}
  \textbf{328} 1658

\bibitem{Wig1955}
Wigner E~P 1955 \emph{Phys. Rev.} \textbf{98} 145

\bibitem{Smi1960}
Smith F~T 1960 \emph{Phys. Rev.} \textbf{118} 349

\bibitem{NagPazFei2011}
Nagele S, Pazourek R, Feist J, Doblhoff-Dier K, Lemell C, T\H{o}k\'{e}si K and
  Burgd\"{o}rfer J 2011 \emph{J. Phys. B} \textbf{44} 081001

\bibitem{KluDahGis2011}
Kl\"{u}nder K, Dahlstr\"{o}m J~M, Gisselbrecht M, Fordell T \emph{et~al.} 2011
  \emph{Phys. Rev. Lett.} \textbf{106} 143002

\bibitem{Schafer2009}
Schafer K~J 2009 in Brabec T (ed.), \emph{Strong Field Laser Physics} (New
  York: Springer) vol. 134 of \emph{Springer Series in Optical Sciences}
  chap.~6, pp. 111--145

\bibitem{MauJohMan2008}
Mauritsson J, Johnsson P, Mansten E, Swoboda M, Ruchon T, L'Huillier A and
  Schafer K~J 2008 \emph{Phys. Rev. Lett.} \textbf{100} 073003

\bibitem{TonChu1997}
Tong X~M and Chu S~I 1997 \emph{Chemical Physics} \textbf{217} 119

\bibitem{deCNus2002}
de~Carvalho C~A~A and Nussenzveig H~M 2002 \emph{Physics Reports} \textbf{364}
  83

\bibitem{SmiSpaIva2006a}
Smirnova O, Spanner M and Ivanov M~Y 2006 \emph{J. Phys. B} \textbf{39} S323

\bibitem{ZhaThu2010}
Zhang C~H and Thumm U 2010 \emph{Phys. Rev. A} \textbf{82} 043405

\bibitem{BagMad2010Err}
Baggesen J~C and Madsen L~B 2010 \emph{Phys. Rev. Lett.} \textbf{104} 209903

\end{thebibliography}

\end{document}